\documentclass[aps,pra,amsmath,amssymb,reprint,groupedaddress,showpacs]{revtex4-1}
\usepackage{graphicx}
\usepackage{float}
\usepackage{dcolumn}
\usepackage{bm}
\newcommand{\n}{\text{-}}

\newcommand{\I}{i}
\newcommand{\hh}{\scriptscriptstyle H\!H\!}
\newcommand{\vv}{\scriptscriptstyle V\!V\!}
\newcommand{\hv}{\scriptscriptstyle H\!V\!}
\newcommand{\vh}{\scriptscriptstyle V\!H\!}
\newcommand{\sss}\scriptscriptstyle
\newcommand{\s}\scriptstyle
\begin{document}

\title{A nonlocal polarization interferometer for entanglement detection}
\author{Brian P. Williams}
\affiliation{Department of Physics and Astronomy, University of Tennessee, Knoxville, Tennessee USA 37996}
\email{bpwilliams@gmail.com}
\author{Travis S. Humble}
\email{humblets@ornl.gov}
\author{Warren P. Grice}
\email{gricew@ornl.gov}
\affiliation{Oak Ridge National Laboratory, 1 Bethel Valley Road, Oak Ridge, Tennessee USA 37831-2008}
\date{\today}
\begin{abstract}
We report a nonlocal interferometer capable of detecting entanglement and identifying Bell states statistically. This is possible due to the interferometer's unique correlation dependence on the anti-diagonal elements of the density matrix, which have distinct bounds for separable states and unique values for the four Bell states. The interferometer consists of two spatially separated balanced Mach-Zehnder or Sagnac interferometers that share a polarization entangled source. Correlations between these interferometers exhibit non-local interference, while single photon interference is suppressed. This interferometer also allows for a unique version of the CHSH-Bell test where the local reality is the photon polarization. We present the relevant theory and experimental results.
\end{abstract}
\maketitle
\section*{Introduction}
\begin{figure}[b]
\centering
\includegraphics[width=7cm]{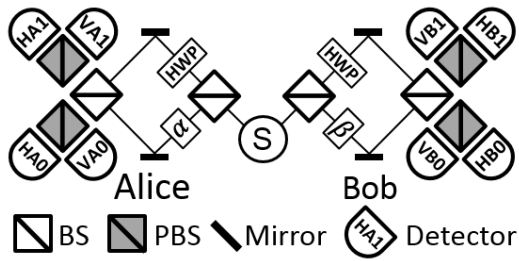}
\caption{The two-photon interferometer is composed of two balanced Mach-Zehnder interferometers sharing a polarization entangled source. Non-local interference effects are observed while single-photon interference is suppressed.\label{BF_Figure}}
\end{figure}
Entanglement enables a variety of proposed quantum information applications \cite{multiphotonreview} such as quantum key distribution \cite{Ekert91}, superdense coding \cite{superdense}, teleportation \cite{teleport}, and quantum computing \cite{qc}. Necessarily, the detection, quantification, and characterization of entanglement is fundamental to its application \cite{qe,entanglement_detection}. One method of ensuring a distributed state is entangled is to perform a Clauser-Horne-Shimony-Holt (CHSH) Bell test \cite{CHSH}, with entanglement detected for Bell parameters $|S|>2$. A second method is to measure the negativity \cite{negativity} of the state, which is an entanglement measure requiring full state tomography. Entanglement may also be revealed via an entanglement witness (EW) \cite{Horodecki1996,terhal2000} which typically requires significantly fewer measurements than full state tomography. The broad EW class includes witness forms of CHSH-Bell tests and negativity tests. In addition to these quantifications and measures there are others \cite{qe,entanglement_detection}. We report a nonlocal polarization interferometer (NPI) that enables entanglement detection and nonlocal statistical Bell state identification. This form of Bell state identification is nonlocal and statistical. Therefore, it is distinct from the local and deterministic measurements used for teleportation and super-dense coding. Instead, nonlocal Bell state identification permits characterizing entanglement between spatially remote subsystems. This is possible due to the NPI's unique correlation dependence on the anti-diagonal elements of the density matrix, which have separable state bounds and unique values for the four Bell states. Additionally, we report an NPI based CHSH-Bell test with the resulting statistics also identifying the Bell state. \\
\indent The balanced Mach-Zehnder implementation of the NPI is illustrated in Fig. \ref{BF_Figure}. Polarization entangled photon pairs are distributed amongst Alice and Bob, each of whom has a balanced Mach-Zehnder interferometer that includes a half-wave plate (HWP) in one path. The HWP is oriented so as to rotate horizontal(vertical) polarization to vertical(horizontal). Upon exiting the interferometers, the photons are directed to polarizing beam splitters (PBSs) monitored by single-photon detectors. Single-photon interference is suppressed by polarization rotation in one path, but two-photon interference remains observable as the phases $\alpha$ and $\beta$ are modulated. Though similar in appearance, the NPI is distinct from the well known Franson interfereometer \cite{Franson89}. Franson's design harnesses time-bin entangled states to demonstrate nonlocal interference while the NPI uses only polarization entanglement. In the remainder of this article we describe the conditions under which correlations are observed, we put bounds on correlations for separable states, we show that the Bell states produce unique NPI signatures, we discuss the NPI version of the CHSH-Bell test, and we present experimental results verifying these predictions using a phase-stable Sagnac version of the NPI.
\section*{Nonlocal Interference}
\begin{figure}[b]
\centering
\includegraphics[width=7cm]{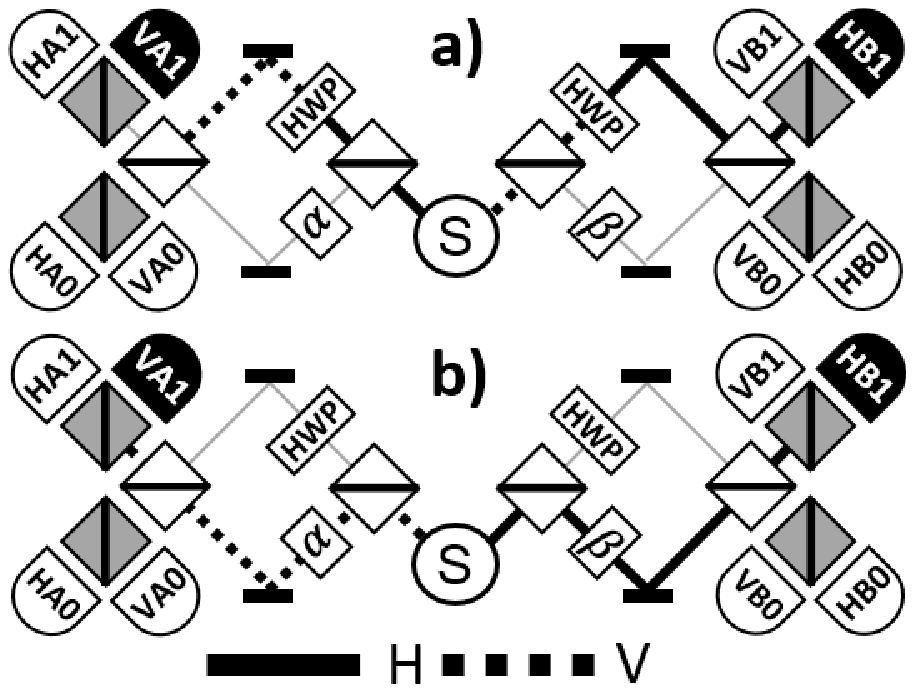}
\caption{Nonlocal interference observed for an orthogonal polarization event VA1HB1 is due to indistinguishable cases a) and b).\label{Franson_different_fig}}
\includegraphics[width=7cm]{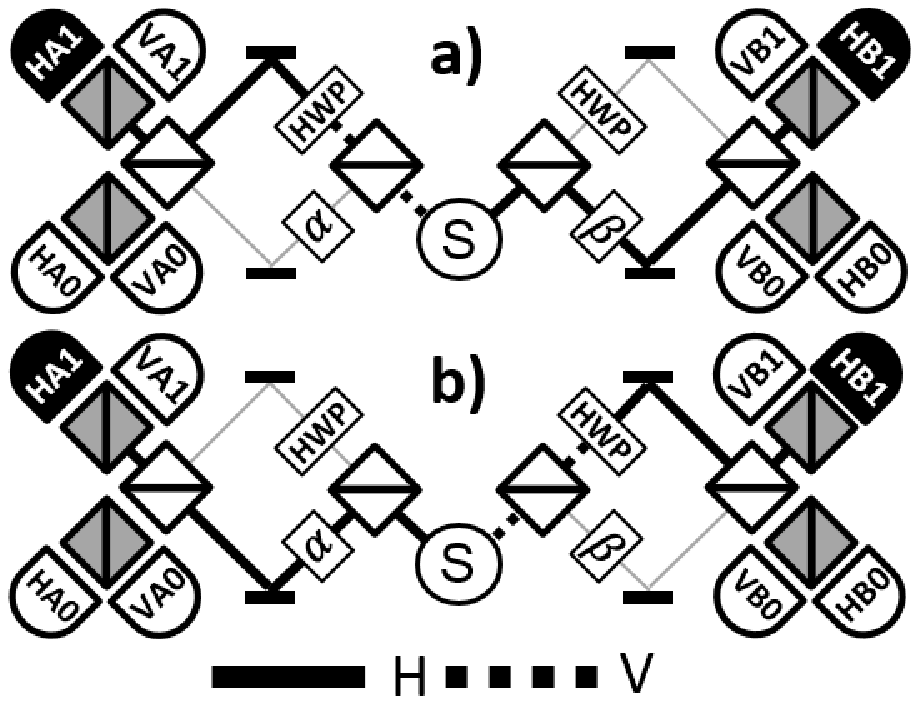}
\caption{Nonlocal interference observed for an identical polarization event HA1HB1 is due to indistinguishable cases a) and b).\label{Franson_same_fig}}\end{figure}
Non-local two-photon interference occurs when Alice and Bob share a polarization entangled source such as the Bell state
\begin{equation}\left|\Psi^+\right\rangle=(1/\sqrt{2})\left(\left|H_A\right\rangle \otimes \left|V_B\right\rangle+\left|V_A\right\rangle\otimes\left| H_B\right\rangle\right) \textrm{.}\label{full pol state}\end{equation}
Given this source, it is straightforward to show that the probability for a single photon to exit any given port of Alice or Bob's Mach-Zehnder interferometer is $1/4$, regardless of phase. That is, no single-photon interference is observed. However, non-local interference is observed in the coincidences between Alice and Bob's detectors. The probability that the signal and idler exit Alice's port $y$ and Bob's port $z$ is
\begin{equation}P_{\substack{jAy\\ sBz}}(\alpha,\beta)\!=\!\!\begin{cases}
   \frac{1}{16}\left\{1 +(\n1)^{z+y}\cos\left[\alpha\!+\!\beta\right]\right\} \!&\! j\!\neq\! s\! \\
   \frac{1}{16}\left\{1 +(\n1)^{z+y}\cos\left[\alpha\!-\!\beta\right]\right\} \!&\! j\!=\!s\!
  \end{cases}
\label{coincidence rate_psi_plus}\end{equation}
where $A$ and $B$ indicate Alice and Bob detectors, respectively, indices $j,s\in\{H,V\}$ indicate the polarization of the detected photons, $y,z\in\{0,1\}$ indicate the detection port, and the phases $\alpha$ and $\beta$ result from the path length mismatch in Alice and Bob's interferometers.\\
\indent Nonlocal interference in the NPI can be understood with the help of Figs. \ref{Franson_different_fig} and \ref{Franson_same_fig}, which show four ways that a coincidence can occur. For the input state given in Eq. \ref{full pol state}, we see in Fig. \ref{Franson_different_fig} that orthogonally polarized photons are detected only if the polarizations of the photons are both rotated by 90 degrees or if they are both left un-rotated, i.e., if both photons take the upper paths or both take the lower paths. These cases are indistinguishable and equally likely, thereby leading to the orthogonal ($j$$\neq$$s$) interference pattern of Eq. \ref{coincidence rate_psi_plus}. Likewise, in Fig. \ref{Franson_same_fig} we see that the photons are detected with identical polarizations only if one travels the upper path and one the lower. Interference between these indistinguishable cases leads to the parallel ($j$$=$$s$) interference pattern in Eq. \ref{coincidence rate_psi_plus}.
\section*{A Phase Stable NPI}
\begin{figure}[b]
\includegraphics[width=8.5cm]{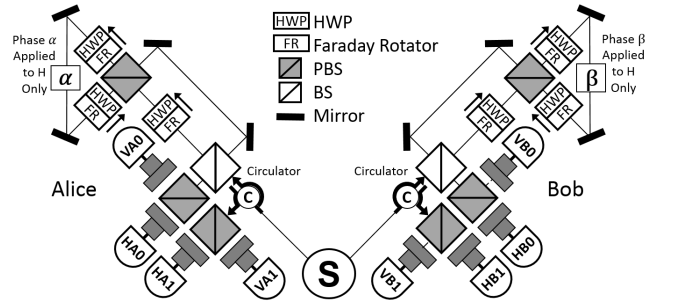}
\caption{Sagnac interferometer setup including directionally dependent phase modulator and dual polarization circulator.\label{sagnac_setup}}
\end{figure}

\indent The Mach-Zehnder version of the NPI is simple and provides insight into the indistinguishable cases leading to interference. However, the Mach-Zehnder interferometer requires active phase stabilization in order to produce stable nonlocal correlations. To avoid this difficulty, we use a Sagnac-based device that allows observation of the same nonlocal interference effects but in a phase-stable configuration. Typically, a fixed Sagnac or ring interferometer's phase cannot be adjusted due to the common path nature of the device. However, our implementation uses directionally dependent phase modulators (DDPM) as well as directionally dependent polarization rotators. The DDPM design is reported in \cite{DDPM} as a polarization independent phase modulator. This design is indeed polarization independent, but it is also directionally dependent, i.e. it matters which port the photon is incident. A schematic of our Sagnac NPI design is given in Fig. \ref{sagnac_setup}. 
Referencing Bob's interferometer in this figure, photons pass through a circulator to the first BS where they randomly choose the reflected, clockwise (CW), path or they choose the transmitted, counter-clockwise (CCW), path. Photons taking the CW path are then incident on the upper PBS input port of the DDPM. The vertical component of these photons travels CW in the DDPM and the horizontal component travels CCW. In either path, the photons encounter Faraday rotator (FR) and half-wave plate (HWP) combinations that are configured to rotate the polarization $90^\circ$ in the direction indicated by the arrow aside the FR-HWP. In the direction opposite these arrows the photon polarization is not changed. The phase modulation $\beta$ is applied only to photons whose polarization is horizontal when they encounter the modulator indicated by $\beta$ in the figure. It is clear that photons taking the reflected (CW) path in Bob's interferometer gain a phase $\beta$, while photons taking the transmitted (CCW) path gain no phase, no matter the polarization. The additional FR-HWP combination in the transmitted path of Bob's interferometer takes on the role of the HWP in the Mach-Zehnder device; It suppresses single-photon interference with a $90^\circ$ polarization rotation for CCW propagating photons. Given that Alice and Bob share the Bell state $\Psi^+$ as in the Mach-Zehnder description, an orthogonal event in the Sagnac NPI could have resulted from the indistinguishable cases that both Alice and Bob's photons are transmitted through the first BS, or both are reflected. Parallel coincidences occur from the indistinguishable cases that Alice's photon takes the transmitted path and Bob's the reflected or vice-versa.
\section*{Separable and Entangled States}
\begin{figure}[H]
\centering
\includegraphics[width=6cm]{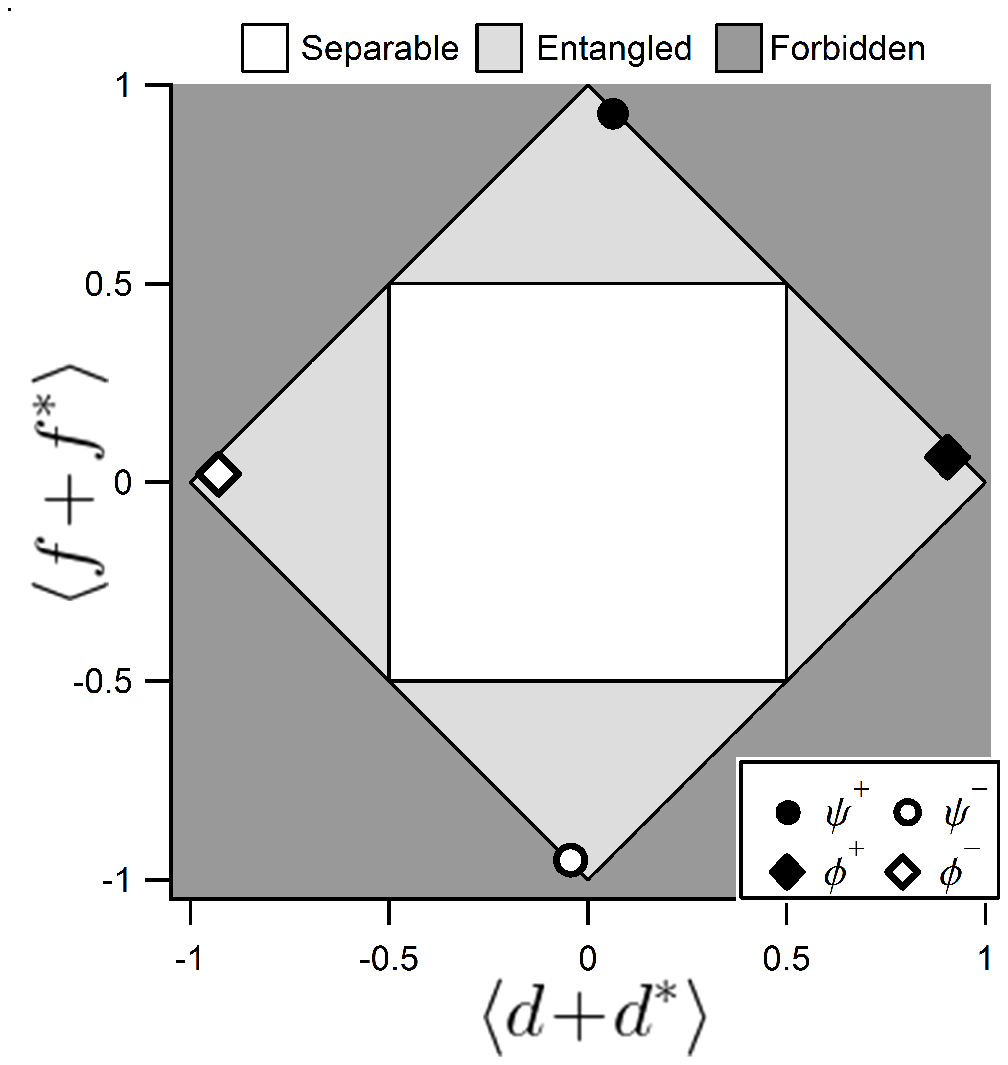}
\caption{Separable and entangled state bounds for parameters $f+f^*$ and $d+d^*$ with corresponding measurement values for the four Bell states indicated by dots and diamonds.\label{Bell_ID}}
\includegraphics[width=6cm]{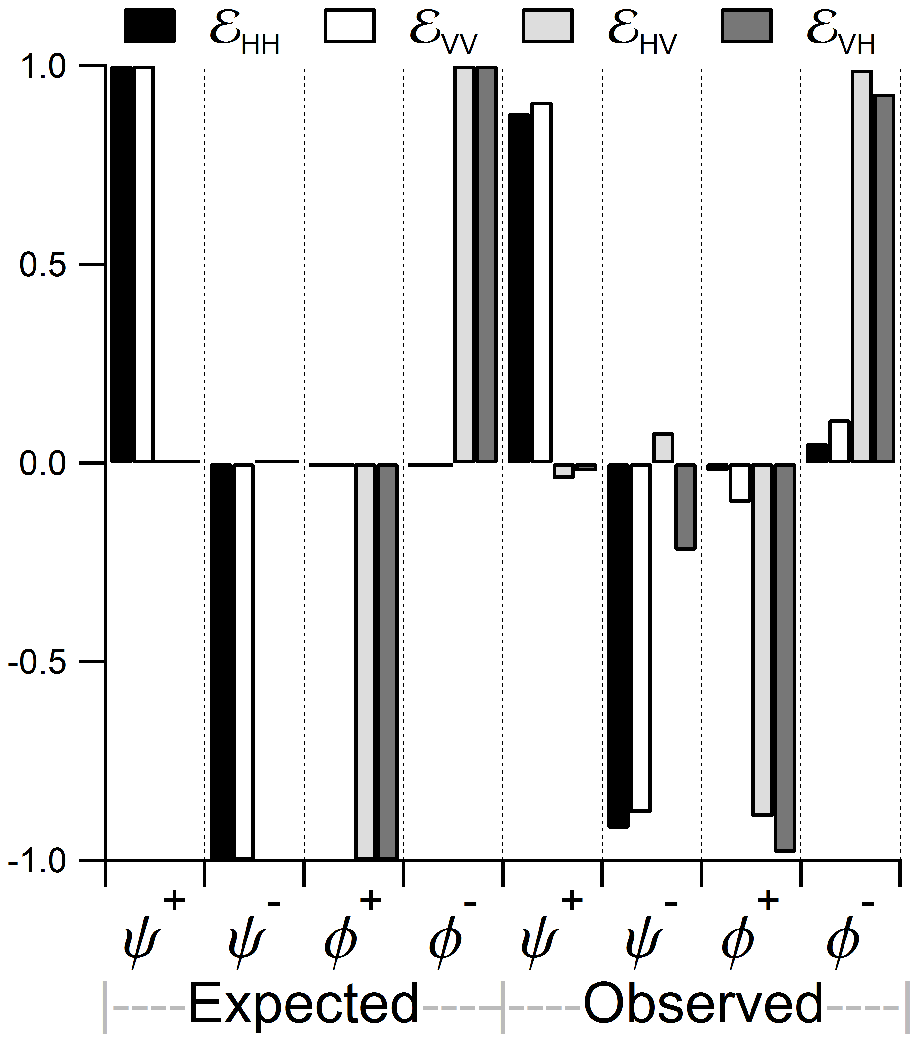}
\caption{Expected and observed values for the correlation coefficients $\mathcal{E}_{\hh}$, $\mathcal{E}_{\vv}$, $\mathcal{E}_{\hv}$, and $\mathcal{E}_{\vh}$ in the standard configuration for each Bell state. Standard deviations for these coefficients are $\leq$ 0.06.\label{bars_standard}}
\includegraphics[width=8cm]{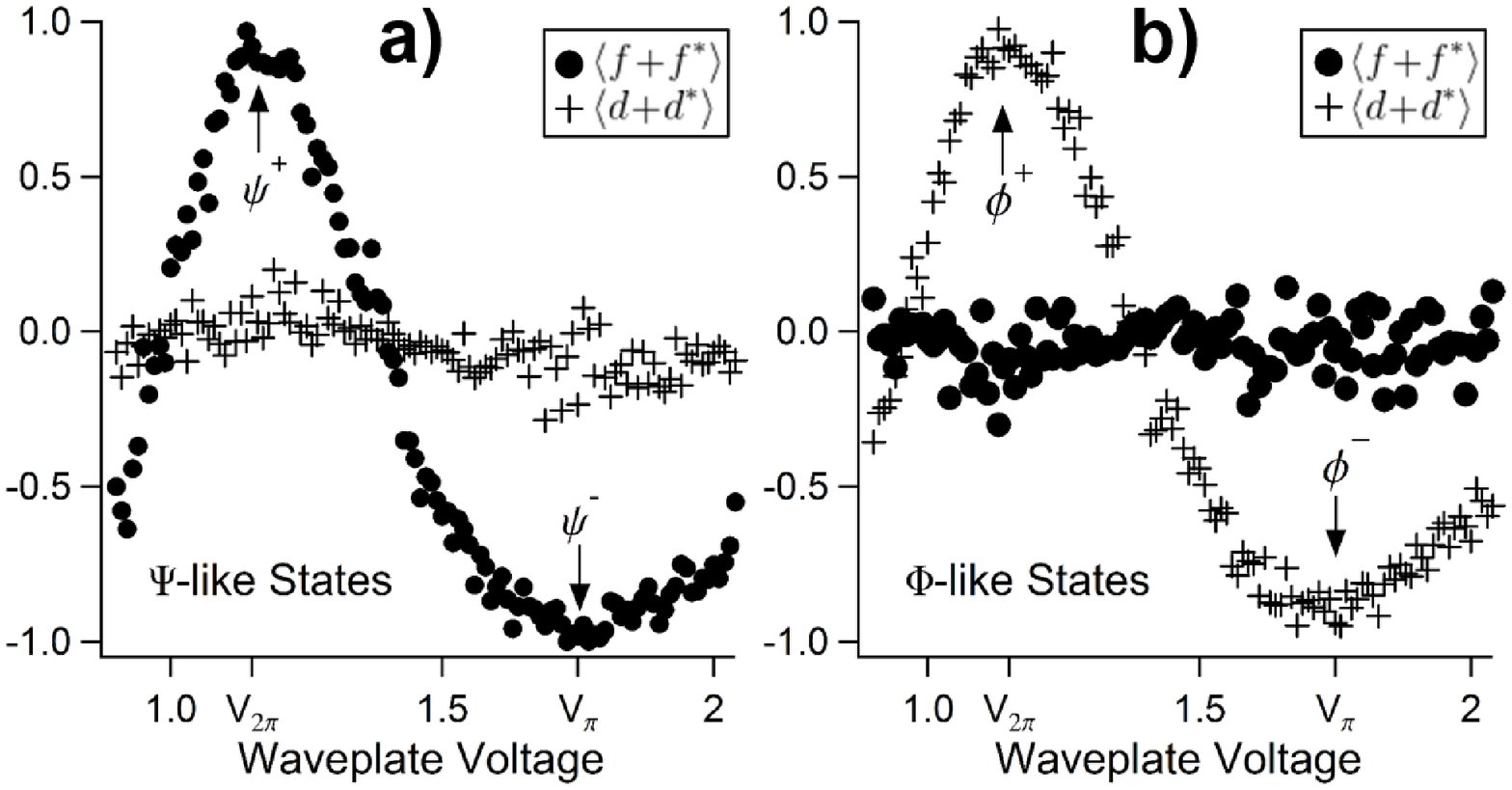}
\caption{a) Variation of phase $\theta$ in state $\psi(\theta)\propto HV+e^{\I\theta}VH$ using a liquid crystal waveplate. b) Variation of phase $\gamma$ in state $\phi(\gamma)\propto HH+e^{\I\gamma}VV$ using a liquid crystal waveplate.\label{unrotated}}
\end{figure}

Assume Alice and Bob share a two-photon state with a density matrix in the polarization basis\begin{equation}\rho=\left(
  \begin{array}{cccc}
    a & b & c & d \\
    b^* & e & f & g \\
    c^* & f^* & h & j \\
    d^* & g^* & j^* & k \end{array}
\right)\textrm{.}\label{generic}\end{equation}The NPI includes two input ports per interferometer, with $H$ and $V$ polarization components. The two-photon state is distributed through one input port of each interferometer, while the remaining port is vacuum, see Fig. \ref{BF_Figure} or \ref{sagnac_setup}. The 16x16 density matrix including all of these components is
\begin{equation}\rho^{'}=\left(
  \begin{array}{ccccccccc}
    a & b & 0 & 0 & c & d & 0&\cdots&0\\
    b^* & e & 0 & 0 & f & g & 0&\cdots&0\\
    0 & 0 & 0 & 0 & 0 & 0 & 0&\cdots&0\\
    0 & 0 & 0 & 0 & 0 & 0 & 0&\cdots&0\\
    c^* & f^* & 0 & 0 & h & k & 0&\cdots&0\\
    d^* & g^* & 0 & 0 & k^* & l & 0&\cdots&0 \\
    0 & 0 & 0 & 0 & 0 & 0 & 0&\cdots&0 \\
    \vdots & \vdots & \vdots & \vdots & \vdots & \vdots & \vdots&\ddots& \\
    0 & 0 & 0 & 0 & 0 & 0 & 0&  &0   \end{array}
\right)\textrm{.}\nonumber\end{equation}
where the zero elements include vacuum components.\\
\indent The operator representing either Alice or Bob's Sagnac interferometer is
\begin{equation}M(\phi)=\left(B\otimes I\right)\cdot\left(e^{\I \phi}Z \oplus X\right)\cdot\left(B\otimes I\right)\nonumber\end{equation}
where $\phi\in\{\alpha,\beta\}$,
\begin{align} I=&\left( \begin{array}{ccc}
1 & 0  \\
0 & 1  \end{array} \right),
\quad
B=\frac{1}{\sqrt{2}}\left( \begin{array}{ccc}
\I & 1  \\
1 & \I  \end{array}\right),\nonumber \\
X=&\left( \begin{array}{ccc}
0 & 1  \\
1 & 0  \end{array} \right),
\quad \textrm{and} \quad
Z=\left( \begin{array}{ccc}
1 & 0  \\
0 & \n1  \end{array} \right).\nonumber
\end{align}
The operator $\otimes$ indicates the Kronecker product and $\oplus$ indicates the direct sum. The operation $B\otimes I$ expands the BS operation from a single component at each input and output port to include both horizontal and vertical polarization components. The operation $e^{\I \phi}Z \oplus X$ represents operations for each path in an interferometer. Photons taking the reflected path in either interferometer have phase modulation $\phi$ applied. Vertical photons taking the reflected path gain an additional $\pi$ phase resultant from the subtleties of the FR-HWP polarization rotation. Photons taking the transmitted path have their polarization rotated $90^\circ$ but gain no phase. The operator for the Mach-Zehnder version of the NPI is found by replacing $Z$ with $I$. This subtle difference changes the form of many of the equations to come, though the same information is extractable from either device. We will use the Sagnac NPI operator, since our experimental results were taken with this device.\\
\indent The final density matrix post interferometers is
\begin{equation}\rho^{''}(\alpha,\beta)=U(\alpha,\beta) \rho^{'} U^{\dagger}(\alpha,\beta)\nonumber\end{equation}
where $U(\alpha,\beta)=M(\alpha)\otimes M(\beta)$. The probability of a coincidence for each combination of Alice and Bob's detectors is given by the diagonal elements of $\rho^{''}(\alpha,\beta)$,
\begin{align}P_{\substack{HA0\\HB0}}(\alpha,\beta)&=\rho_{11}^{''}(\alpha,\beta) \quad P_{\substack{HA0\\VB0}}(\alpha,\beta)=\rho_{22}^{''}(\alpha,\beta)\nonumber\\
P_{\substack{HA0\\HB1}}(\alpha,\beta)&=\rho_{33}^{''}(\alpha,\beta) \quad P_{\substack{HA0\\VB1}}(\alpha,\beta)=\rho_{44}^{''}(\alpha,\beta)\nonumber \\
P_{\substack{VA0\\HB0}}(\alpha,\beta)&=\rho_{55}^{''}(\alpha,\beta) \quad P_{\substack{VA0\\VB0}}(\alpha,\beta)=\rho_{66}''(\alpha,\beta)\nonumber\\
P_{\substack{VA0\\HB1}}(\alpha,\beta)&=\rho_{77}^{''}(\alpha,\beta) \quad P_{\substack{VA0\\VB1}}(\alpha,\beta)=\rho_{88}^{''}(\alpha,\beta)\nonumber \\
P_{\substack{HA1\\HB0}}(\alpha,\beta)&=\rho_{99}^{''}(\alpha,\beta) \quad P_{\substack{HA1\\VB0}}(\alpha,\beta)=\rho_{\substack{10\\10}}^{''}(\alpha,\beta)\nonumber\\
P_{\substack{HA1\\HB1}}(\alpha,\beta)&=\rho_{\substack{11\\11}}^{''}(\alpha,\beta) \quad P_{\substack{HA1\\VB1}}(\alpha,\beta)=\rho_{\substack{12\\12}}^{''}(\alpha,\beta)\nonumber \\
P_{\substack{VA1\\HB0}}(\alpha,\beta)&=\rho_{\substack{13\\13}}^{''}(\alpha,\beta) \quad P_{\substack{VA1\\VB0}}(\alpha,\beta)=\rho_{\substack{14\\14}}^{''}(\alpha,\beta)\nonumber\\
P_{\substack{VA1\\HB1}}(\alpha,\beta)&=\rho_{\substack{15\\15}}^{''}(\alpha,\beta) \quad P_{\substack{VA1\\VB1}}(\alpha,\beta)=\rho_{\substack{16\\16}}^{''}(\alpha,\beta)\nonumber
\textrm{.}\end{align}
When the source is a Bell state,
\begin{align}\left|\Phi^\pm\right\rangle&=(1/\sqrt{2})\left(\left|H_A\right\rangle \otimes \left|H_B\right\rangle\pm\left|V_A\right\rangle\otimes\left| V_B\right\rangle\right) \quad \textrm{or} \nonumber\\
\left|\Psi^\pm\right\rangle&=(1/\sqrt{2})\left(\left|H_A\right\rangle \otimes \left|V_B\right\rangle\pm\left|V_A\right\rangle\otimes\left| H_B\right\rangle\right)\textrm{,}\nonumber
\end{align}
 we find the probabilities
\begin{equation}P_{\substack{jAy\\ sBz}}(\alpha,\beta)\!=\!\!\begin{cases}
   \frac{1}{16}\left\{1\!+\ell(\n1)^{z+y}\cos\left[\alpha\!+\!m\beta\right]\right\} \!&\! j\!\neq\! s\! \\
   \frac{1}{16}\left\{1\!+\ell(\n1)^{z+y}\cos\left[\alpha\!-\!m\beta\right]\right\} \!&\! j\!=\!s\!
  \end{cases}
\nonumber\end{equation}
where $\ell,m$ values for each Bell state are $\Psi^+$:\{1,1\},$\Psi^-$:\{-1,1\},$\Phi^+$:\{1,-1\}, and $\Phi^-$:\{-1,-1\}.\\
\indent The probability of coincidence for any given port combination depends on the density matrix elements given in Eq. \ref{generic}. This dependence varies with the phases $\alpha$ and $\beta$. However, the case $\alpha$=$\beta$= $\pi/4$ is particularly interesting. It is this case that the remainder of this paper will focus on. We refer to the NPI configured with $\alpha$=$\beta$= $\pi/4$ as NPI$_{\pi \!/\! _4}$, the standard configuration. With these settings, it is straightforward to show that for the general density matrix $\rho$ given in Eq. \ref{generic} that\begin{widetext}
\begin{equation}P_{\substack{jAy\\ sBz}}\!\!\left(\frac{\pi}{4},\frac{\pi}{4}\right)\!=\!\!\begin{cases}
   \frac{1}{16}\left[ 1-(\n1)^{y+z}\left\{d+d^*+i(\delta_{jH}-\delta_{jV})(f-f^*)\right\}+(\n 1)^y \sigma_{jA}+(\n 1)^z \sigma_{sB} \right] \!&\! j\!\neq\! s\! \\
      \frac{1}{16}\left[ 1+(\n1)^{y+z}\left\{i(\delta_{jH}-\delta_{jV})(d-d^*)+f+f^*\right\}+(\n 1)^y \sigma_{jA}+(\n 1)^z \sigma_{sB} \right] \!&\! j\!=\! s\! \\
  \end{cases}
\nonumber\end{equation}\end{widetext}
where
\begin{align}\sigma_{_{HA}}&=e^{\I\pi/4}\{-c-g\!+\I(c^*+g^*)\}\textrm{,} \nonumber \\
\sigma_{_{V\!A}}&=e^{\I\pi/4}\{c^*+g^*-\I(c+g)\}\textrm{,} \nonumber \\
\sigma_{_{H\!B}}&=e^{\I\pi/4}\{-b-k+\I(b^*+k^*)\}\textrm{, and} \nonumber \\
\sigma_{_{V\!B}}&=e^{\I\pi/4}\{b^*+k^*-\I(b+k)\}\nonumber
\end{align}
are proportional to the marginal coherences, i.e. single-photon interference.
Defining the polarization dependent correlation coefficient as
\begin{equation}
\mathcal{E}_{\!js}\equiv\!\!\frac{\!P_{\!\substack{ \sss j\!A0\\ \sss s\!B0}}(\frac{\pi}{4}\!,\!\frac{\pi}{4})\!+\!P_{\!\substack{ \sss j\!A1\\ \sss s\!B1}}( \frac{\pi}{4}\!,\!\frac{\pi}{4})\!-\!P_{\!\substack{\sss j\!A0\\ \sss s\!B1}}( \frac{\pi}{4}\!,\!\frac{\pi}{4})\!-\!P_{\!\substack{\sss j\!A1\\ \sss s\!B0}}( \frac{\pi}{4}\!,\!\frac{\pi}{4})\!}{\!P_{\!\substack{ \sss j\!A0\\ \sss s\!B0}}( \frac{\pi}{4}\!,\!\frac{\pi}{4})\!+\!P_{\!\substack{ \sss j\!A1\\ \sss s\!B1}}( \frac{\pi}{4}\!,\!\frac{\pi}{4})\!+\!P_{\!\substack{\sss j\!A0\\ \sss s\!B1}}( \frac{\pi}{4}\!,\!\frac{\pi}{4})\!+\!P_{\!\substack{\sss j\!A1\\ \sss s\!B0}}( \frac{\pi}{4}\!,\!\frac{\pi}{4})\!}\textrm{,}\label{coeff}\end{equation}
we find the real-valued coefficients
\begin{align}
\mathcal{E}_{\hh}&=f+f^*-\I(d-d^*)\textrm{,} \label{hh0}\\
\mathcal{E}_{\vv}&=f+f^*+\I(d-d^*)\textrm{,} \label{vv0}\\
\mathcal{E}_{\hv}&=-d-d^*-\I(f-f^*)\textrm{, and}\label{hv0} \\
\mathcal{E}_{\vh}&=-d-d^*+\I(f-f^*) \label{vh0}\end{align}
where $d,d^*,f,$ and $f^*$ are the anti-diagonal elements of the density matrix $\rho$ given in Eq. \ref{generic}. The correlation coefficients  have values $-1\leq \mathcal{E}\leq 1$, with 1(-1) indicating perfect correlation(anti-correlation). Clearly, we have the resulting relations
\begin{align}f+f^*&= \frac{\phantom{-}\mathcal{E}_{\hh}+\mathcal{E}_{\vv}\!}{2}\textrm{ and}\label{f+f}\\
d+d^*&= \frac{-\mathcal{E}_{\hv}-\mathcal{E}_{\vh}}{2}\textrm{.}\label{d+d}\end{align}
These relations indicate that parallel correlations, $HH$ and $VV$, are proportional to $f+f^*$. Similarly, orthogonal correlations, $HV$ and $VH$, are proportional to $d+d^*$. Referencing Table \ref{elements}, Fig. \ref{Bell_ID}, and \ref{bars_standard} we see that each of the Bell states has a unique correlation signature in the NPI. As an example, when measurements are made on the Bell state $\Psi^+$ we expect no correlation for orthogonal events and perfect correlation for parallel events.

Additionally, it should be clear that the correlation coefficients \ref{hh0}, \ref{vv0}, \ref{hv0}, and \ref{vh0} may also be used to identify the ``shifted" Bell states
\begin{align}\left|\Phi^\pm_s\right\rangle&=(1/\sqrt{2})\left(\left|H_A\right\rangle \otimes \left|H_B\right\rangle\pm\I\left|V_A\right\rangle\otimes\left| V_B\right\rangle\right) \\
\left|\Psi^\pm_s\right\rangle&=(1/\sqrt{2})\left(\left|H_A\right\rangle \otimes \left|V_B\right\rangle\pm\I\left|V_A\right\rangle\otimes\left| H_B\right\rangle\right)\textrm{.}
\end{align}
Thus, eight maximally entangled states may be uniquely identified, statistically, in the NPI.
\begin{table}[H]
    \centering
    \begin{tabular}{| c | c | c | c | c |}
    \hline
                         &$\Psi^+ $&$ \Psi^-$ &$\Phi^+$ &$\Phi^-$ \\ \hline
    $d+d^*$&0&0&1&-1 \\ \hline
    $f+f^*$&1&-1&0&0 \\ \hline
\end{tabular}\caption{Bell state signatures for $f+f^*$ and $d+d^*$.\label{elements}}\end{table}

For any state, $|f+f^*|\leq1$ and $|d+d^*|\leq 1$. However, if we consider the density matrix for a separable pure state
\begin{equation}\rho_{A}\otimes\rho_B=\left|A\right\rangle\left\langle A\right|\otimes\left|B\right\rangle\left\langle B\right|\nonumber\end{equation}
where
\begin{equation}
\left|A\right\rangle \!=\!\left(\!
  \begin{array}{c}
    \sin(a) \\
    \cos(a)e^{i\theta_A} \\  \end{array}\!\right) \quad\textrm{and}\quad \left|B\right\rangle\!=\!\left(\!
  \begin{array}{c}
    \sin(b) \\
    \cos(b)e^{i\theta_B} \\
  \end{array}\!\right)\textrm{,}\nonumber\end{equation}
we find
\begin{align}f+f^*&=(1/2)\sin(2a)\sin(2b)\cos(\theta_A-\theta_B) \nonumber\\
d+d^*&=(1/2)\sin(2a)\sin(2b)\cos(\theta_A+\theta_A)\nonumber\end{align}
which requires
\begin{equation}\left|f+f^*\right|\leq\frac{1}{2}\quad \textrm{and}\quad \left|d+d^*\right|\leq\frac{1}{2}\textrm{.}\label{bounds}\end{equation}
These inequalities also hold for any separable mixed state of the form
\begin{equation}\rho_{mix}=\sum_\lambda p_\lambda \rho_{A}^{\lambda}\otimes\rho_{B}^{\lambda}\textrm{,}\nonumber\end{equation}
since, in this case,
\begin{align}f+f^*&=\sum_\lambda p_\lambda \left(f_\lambda+f^*_\lambda\right) \quad \textrm{and} \nonumber \\
 d+d^*&=\sum_\lambda p_\lambda \left(d_\lambda+d^*_\lambda\right)\textrm{.}\nonumber\end{align}
Thus, the conditions $|f+f^*|>1/2$ or $|d+d^*|>1/2$ are required for an entangled state.

Knowledge of $f+f^*$ and $d+d^*$ also determine the minimum Bell state fidelities. The fidelities or overlap of the generic density matrix from Eq. \ref{generic} with each Bell state are
\begin{align}F_{\Phi^\pm}&=\left\langle\Phi^\pm\right|\rho\left|\Phi^\pm\right\rangle=\left(a+k\pm[d+d^*]\right)/2\label{fid_phi} \;\;\;\textrm{and}\\
F_{\Psi^\pm}&=\left\langle\Psi^\pm\right|\rho\left|\Psi^\pm\right\rangle=\left(e+h\pm[f+f^*]\right)/2\label{fid_psi}\textrm{.}\end{align}
Since all density matrices must be positive semi-definite, $\left\langle \phi\right|\rho\left|\phi\right\rangle \geq 0$, Eq. \ref{fid_phi} and \ref{fid_psi} require
\begin{equation}a+k \geq \left|d+d^*\right|\quad\quad\textrm{and}\quad\quad e+h \geq \left|f+f^*\right|\textrm{.}\nonumber\end{equation}
These inequalities lead to the the minimum fidelity values
\begin{align}F_{\psi^+}&\geq(\left|f+f^*\right|+f+f^*)/2\textrm{,}\nonumber\\
F_{\psi^-}&\geq(\left|f+f^*\right|-f-f^*)/2\textrm{,}\nonumber\\
F_{\phi^+}&\geq(\left|d+d^*\right|+d+d^*)/2\textrm{, and}\nonumber\\
F_{\phi^-}&\geq(\left|d+d^*\right|-d-d^*)/2\textrm{.}\nonumber\end{align}
Only one of these can exceed $1/2$ for a given state.

Experimentally, we determine the expectation value of the correlation coefficient given in Eq. \ref{coeff} as
\begin{equation}
\left\langle\mathcal{E}_{\!js}\right\rangle\!=\!\!\frac{\!\mathcal{C}_{\substack{ \sss j\!A0\\ \sss s\!B0}}(\frac{\pi}{4}\!,\!\frac{\pi}{4})\!+\!\mathcal{C}_{\substack{ \sss j\!A1\\ \sss s\!B1}}( \frac{\pi}{4}\!,\!\frac{\pi}{4})\!-\!\mathcal{C}_{\substack{\sss j\!A0\\ \sss s\!B1}}( \frac{\pi}{4}\!,\!\frac{\pi}{4})\!-\!\mathcal{C}_{\substack{\sss j\!A1\\ \sss s\!B0}}( \frac{\pi}{4}\!,\!\frac{\pi}{4})\!}{\!\mathcal{C}_{\substack{ \sss j\!A0\\ \sss s\!B0}}( \frac{\pi}{4}\!,\!\frac{\pi}{4})\!+\!\mathcal{C}_{\substack{ \sss j\!A1\\ \sss s\!B1}}( \frac{\pi}{4}\!,\!\frac{\pi}{4})\!+\!\mathcal{C}_{\substack{\sss j\!A0\\ \sss s\!B1}}( \frac{\pi}{4}\!,\!\frac{\pi}{4})\!+\!\mathcal{C}_{\substack{\sss j\!A1\\ \sss s\!B0}}( \frac{\pi}{4}\!,\!\frac{\pi}{4})\!}\textrm{,}\nonumber\end{equation}
where $\mathcal{C}_{\substack{ \sss j\!Ay\\ \sss s\!Bz}}( \frac{\pi}{4},\frac{\pi}{4})$ are accidental corrected and normalized coincidence counts for detector combinations $jAy$ and $sBz$. The experimental measurements of $f+f^*$ and $d+d^*$ made on many copies of an identical state are
\begin{align}\left\langle f+f^* \right\rangle &= \frac{\phantom{-}\left\langle\mathcal{E}_{\hh}\right\rangle+\left\langle\mathcal{E}_{\vv}\right\rangle}{2}\textrm{ and}\nonumber\\
\left\langle d+d^* \right\rangle&= \frac{-\left\langle\mathcal{E}_{\hv}\right\rangle-\left\langle\mathcal{E}_{\vh}\right\rangle}{2}\textrm{.}\nonumber\end{align}

Entanglement is detected when sufficient experimental statistics are gathered to indicate that
\begin{align}|\left\langle f+f^* \right\rangle|&>1/2 \textrm{ or}\label{exp1} \\
|\left\langle d+d^* \right\rangle|&>1/2\textrm{.}\label{exp2}\end{align}
Maximally entangled states will have experimental values  $|\left\langle f+f^* \right\rangle|\rightarrow 1$ or $|\left\langle d+d^* \right\rangle|\rightarrow 1$.
For any experiment in which the source is static, unchanging, these simple frequency based statistics will hold, and confirmation of Eq. \ref{exp1} or \ref{exp2} will indicate an entangled state with high confidence. Experimental results for each Bell state are plotted in Fig. \ref{Bell_ID} along with  a graphical depiction of the bounds given for separable and entangled states. These results with standard deviations are also given in Table \ref{results}.
We have also given the expected and observed values for the correlation coefficients $\mathcal{E}_{\hh}$, $\mathcal{E}_{\vv}$, $\mathcal{E}_{\hv}$, and $\mathcal{E}_{\vh}$ for each Bell state in the standard configuration in Fig. \ref{bars_standard}. Each of these figures graphically depicts each Bell state's unique correlation signature.

To further illustrate that correlations are directly linked to $f+f^*$ and $d+d^*$, we vary these values using a phase modulator and experimentally determine their value. Results for ``$\psi$"-like states $\psi(\theta)\propto HV+e^{\I\theta}VH$ are given in Fig. \ref{unrotated} (a), where $\cos\theta=f+f^*$ and $\theta$ varies with voltage. Similarly, the results for ``$\phi$"-like states $\phi(\gamma)\propto HH+e^{\I\gamma}VV$ are given in Fig. \ref{unrotated} (b), where $\cos\gamma=d+d^*$ and $\gamma$ varies with voltage. The voltages $V_{2\pi}$ and $V_\pi$ are associated with ``+" and ``-" Bell states, respectively. The phase dependence on voltage is nonlinear but approaches linearity in the region between $V_{2\pi}$=$1.15V$ and $V_\pi$=$1.75V$.

\section*{CHSH with the NPI}
\begin{figure}[b]
\centering
\includegraphics[width=6cm]{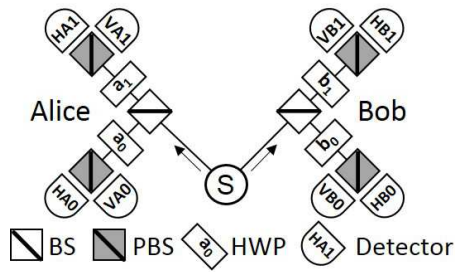}
\caption{A typical CHSH Bell test. An entangled photon pair is shared between Alice and Bob whose local measurement settings or local realities $a_0,a_1,b_0,$ and $b_1$ are randomly chosen by a symmetric beamsplitter.\label{CHSH_fig}}\end{figure}

CHSH-Bell tests \cite{CHSH} are commonly carried out using a polarization-based experiment as seen in Fig. \ref{CHSH_fig}. For states that obey locality constraints, the Bell parameter $S$ obeys the inequality
\begin{equation}\left|S\right|\!=\left|\!E(a_0,b_0)+E(a_0,b_1)+E(a_1,b_0)-E(a_1,b_1)\right|\!\leq\!2\nonumber\end{equation}
where $a_0,a_1,b_0,$ and $b_1$ are local realities such as polarization rotation, and the correlation coefficient is
\begin{equation} E(a,b)=\!P_{\!\hh}(a,b)\!+\!P_{\!\vv}(a,b)\!-\!P_{\!\hv}(a,b)\!-\!P_{\!\vh}(a,b) \nonumber\end{equation}
where $P_{js}(a,b)$ is the probability a coincidence between Alice's $j$ polarization detector and Bob's $s$ polarization detector given the local realities $a$ and $b$ for Alice and Bob, respectively.
The experimental estimate of the correlation coefficient in the CHSH-Bell test is
\begin{equation}\langle E(a,b)\rangle\!=\!\frac{\mathcal{C}_{\hh}({a,b})\!+\!\mathcal{C}_{\vv}(a,b)\!-\!\mathcal{C}_{\hv}(a,b)\!-\!\mathcal{C}_{\vh}(a,b)}{\mathcal{C}_{\hh}(a,b)\!+\!\mathcal{C}_{\vv}(a,b)\!+\!\mathcal{C}_{\hv}(a,b)\!+\!\mathcal{C}_{\vh}(a,b)}\nonumber\end{equation}
with coincidence counts $\mathcal{C}_{js}(a,b)$.

Since the polarization of any photon exiting the NPI is random, the final photon polarization represents a ``local reality". Therefore, a unique CHSH-Bell test may be performed based on the four random photon polarization outcomes $HH$, $VV$, $HV$, or $VH$. To maximize inequality breaking, Bob applies a $\pi/4$ phase to his vertical photon prior to its entry into his interferometer. We call this configuration  NPI$_{\pi\!/\!_4}^{\textrm{Bell}}$. In this case, the correlation coefficients are
\begin{align}
\mathcal{E}_{\hh}^{'} &=2^{\;\n\frac{1}{2}}\left(f\!+\!f^*\!-\!\I(f\!-\!f^*)\!+\!d\!+\!d^*\!+\!\I(d\!-\!d^*)\right) \label{hh1}\\
\mathcal{E}_{\vv}^{'} &=2^{\;\n\frac{1}{2}}\left(f\!+\!f^*\!-\!\I(f\!-\!f^*)\!-\!d\!-\!d^*\!-\!\I(d\!-\!d^*)\right) \label{vv1}\\
\mathcal{E}_{\hv}^{'} &=2^{\;\n\frac{1}{2}}\left(-f\!-\!f^*\!-\!\I(f\!-\!f^*)\!-\!d\!-\!d^*\!+\!\I(d\!-\!d^*)\right)\label{hv1} \\
\mathcal{E}_{\vh}^{'} &=2^{\;\n\frac{1}{2}}\left(\!f\!+\!f^*\!+\!\I(f\!-\!f^*)\!-\!d\!-\!d^*\!+\!\I(d\!-\!d^*)\right)\textrm{.}\label{vh1}\end{align}
We define the two Bell parameters
\begin{align}
S_{\psi}\!&\equiv\!\mathcal{E}_{\hh}^{'}\!+\mathcal{E}_{\vv}^{'} \!-\mathcal{E}_{\hv}^{'}\!+\mathcal{E}_{\vh}^{'}\!=2\sqrt{2\!}\left(f\!+\!f^*\right) \textrm{ and}\label{Spsi} \\
S_{\phi}\!&\equiv\!\mathcal{E}_{\hh}^{'}\!-\mathcal{E}_{\vv}^{'}\!-\mathcal{E}_{\hv}^{'}\!-\mathcal{E}_{\vh}^{'}\!=2\sqrt{2\!}\left( d+\!d^*\right) \textrm{.}\!\!\label{Sphi}\end{align}
The Bell parameter $S$ is proportional to the anti-diagonal elements of the density matrix, similar to the results in the last section. As indicated by Eq. \ref{bounds}, these Bell parameters have a separable state bound of $|S|\leq \sqrt{2}$ based on quantum mechanics, not on arguments of locality. Clearly, the $S$ parameters' dependence on $f+f^*$ and $d+d^*$ enable Bell state identification, as was possible in the standard configuration NPI$_{\pi\!/\!_4}$. As is in the standard configuration, Bell parameters can also be defined for each of the shifted Bell states.

\begin{figure}[H]
\centering
\includegraphics[width=6cm]{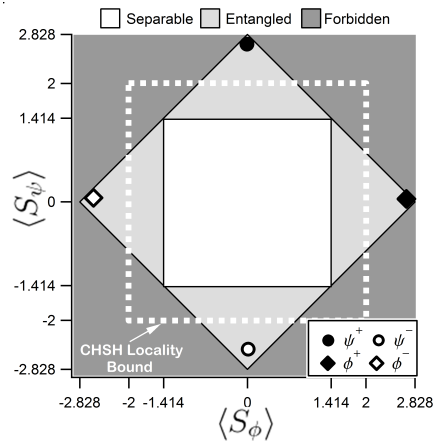}
\caption{Separable and entangled state bounds for parameters $\mathcal{S}_\psi$ and $\mathcal{S}_\phi$ with corresponding measurement values for the four prepared Bell states indicated by dots and diamonds.\label{CHSH_ID}}
\end{figure}

\begin{figure}[H]
\centering
\includegraphics[width=6cm]{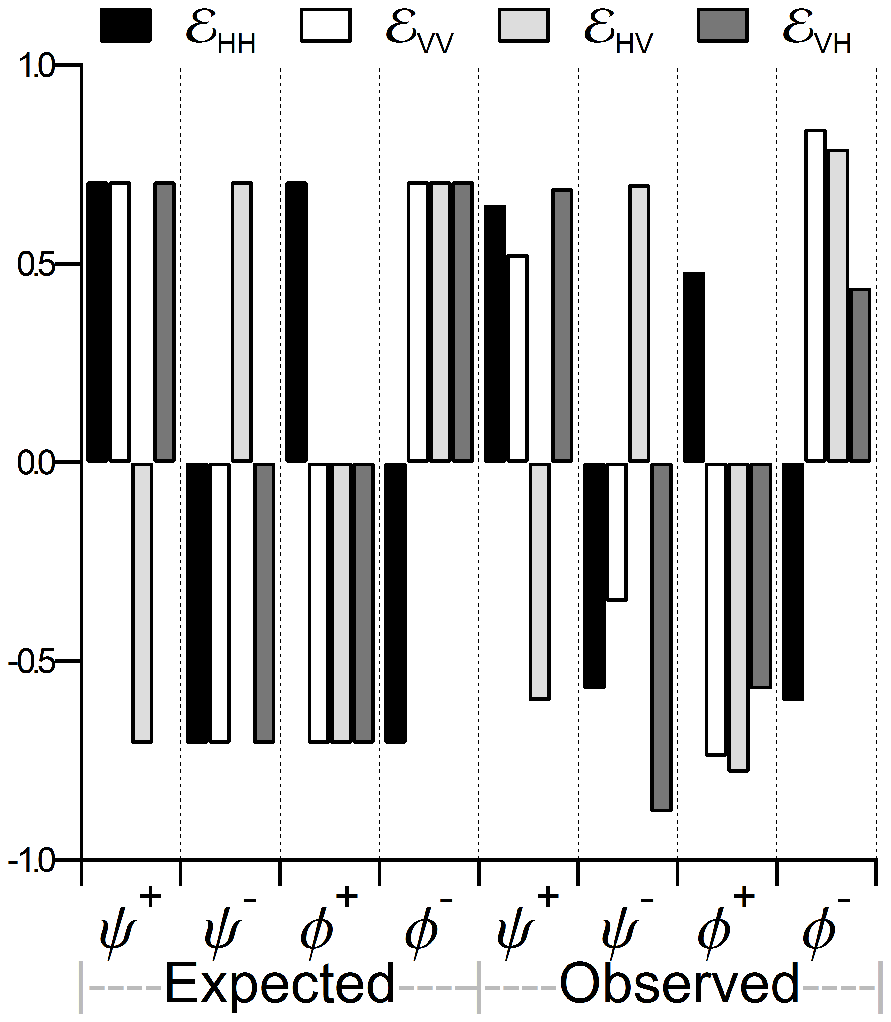}
\caption{Expected and observed values for the correlation coefficients $\mathcal{E}_{\hh}$, $\mathcal{E}_{\vv}$, $\mathcal{E}_{\hv}$, and $\mathcal{E}_{\vh}$ in the CHSH configuration for each Bell state. Standard deviations for these coefficients are $\leq$ 0.06. \label{bars_CHSH}}\end{figure}

The experimental measurements
\begin{align}
\left\langle S_{\psi}\right\rangle\!&=\!\langle\mathcal{E}_{\hh}^{'}\rangle\!+\langle\mathcal{E}_{\vv}^{'}\rangle \!-\langle\mathcal{E}_{\hv}^{'}\rangle\!+\langle\mathcal{E}_{\vh}^{'}\rangle\!=2\sqrt{2\!}\left\langle{f\!+\!f^*}\right\rangle  \nonumber\\
\left\langle S_{\phi}\right\rangle\!&=\!\langle\mathcal{E}_{\hh}^{'}\rangle\!-\langle\mathcal{E}_{\vv}^{'}\rangle\!-\langle\mathcal{E}_{\hv}^{'}\rangle\!-\langle\mathcal{E}_{\vh}^{'}\rangle\!=2\sqrt{2\!}\left\langle{d+\!d^*}\right\rangle \nonumber\end{align}
for each Bell state are graphically depicted in Fig. \ref{CHSH_ID} and given with standard deviations in Table \ref{results}. As they should, the Bell parameters $\langle S_\psi\rangle$ and $\langle S_\phi\rangle$ exceed the $|S|\leq 2$ bound for the appropriate Bell states. We have also given the expected and observed values for the correlation coefficients $\mathcal{E}_{\hh}$, $\mathcal{E}_{\vv}$, $\mathcal{E}_{\hv}$, and $\mathcal{E}_{\vh}$ for each Bell state in the CHSH configuration in Fig. \ref{bars_CHSH}. Each of these figures graphically depicts the unique correlation signatures for each Bell state.

\section*{Experimental Setup}
\begin{figure}[b]
\centering
\includegraphics[width=8.0cm]{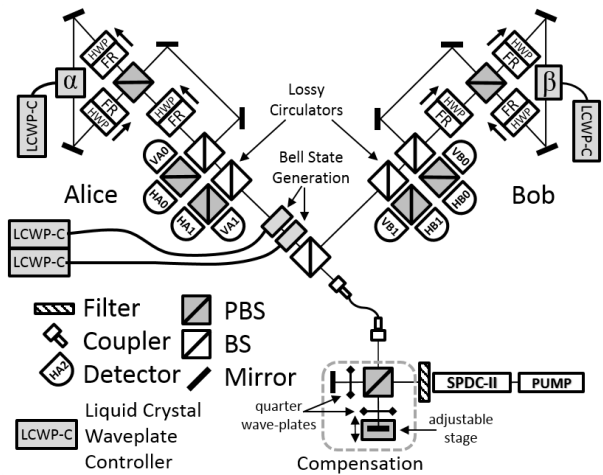}
\caption{The NPI Sagnac experiment includes a polarization entangled source dependent on the spectral indistinguishability of the signal and idler photons. Bell states were generated by polarization rotation and phase modulation in the path to Alice's interferomter.\label{sagnac_experiment}}
\end{figure}
Our experimental results were observed using the apparatus shown in Fig. \ref{sagnac_experiment}. In this experiment 0.9 mW of 405 nm continuous-wave (CW) diode laser light pumps a PPKTP crystal generating approximately $1.4\times10^6 $ Type II signal-idler pairs per second at a wavelength of 810 nm from the spontaneous parametric downconversion process. The signal and idler pass through a compensation system which removes the phase resultant from the polarization-dependent walk-off due to the nonlinear crystal and optical fiber. After passing through the fiber, photons are incident on a BS which produces a polarization entangled shared state when the signal and idler are spectrally indistinguishable \footnote{A non-degenerate signal and idler would destroy the polarization entanglement, since the photon energy would be associated with a specific polarization. This setup also leads to local two-photon interference when the photons both go to Alice or Bob's interferometers. A local effect observed is Hong-Ou-Mandel (HOM) interference which is maximized when the signal and idler are spectrally indistinguishable. The visibility of HOM interference was used to tune the signal and idler indistinguishability.}.
Beamsplitters are used as lossy circulators. Phase modulation in each interferometer is applied using liquid crystal variable wave plates. At each interferometer output port a PBS routes each polarization to separate single-photon detectors through single-mode fiber. In order to determine coincidence rates we time-stamp the detection signal from each single-photon detector into 5 ns time bins using an FPGA \footnote{See www.zedboard.com and www.xillibus.com.}. Eight Perkin-Elmer, now Excelitas, SPCM devices are used to count single-photons. Single photon count rates range from 3-10 kcps and average coincidence rates range from 2-20 cps. Results in Fig. \ref{Bell_ID}, \ref{bars_standard}, \ref{CHSH_ID}, \ref{bars_CHSH}, and Table \ref{results} were generated from 100 sec counts. Results of phase variation, seen in Fig. \ref{unrotated}, were generated from 5 sec counts at each voltage setting.In order to normalize our coincidence counts, we calibrate the NPI with a known unentangled source \footnote{The signal and idler were made spectrally distinguishable by tuning the temperature of the PPKTP crystal. Distinguishability was verified by vanishing local HOM interference, see [14]. This distinguishability destroys the polarization entanglement.} which provides the same flux of photons in each path. This enables determination of the relative efficiency of each detector combination. Due to imperfect optics and experimental shortcomings, single-photon interference is observed with $1\%$ visibility.

Our experimental results qualitatively agree with our theoretical predictions. This can be seen for the standard configuration by comparing Eqs. \ref{hh0}, \ref{vv0}, \ref{hv0}, \ref{vh0}, \ref{f+f}, \ref{d+d}, and Table \ref{elements} with Figs. \ref{Bell_ID}, \ref{bars_standard}, and Table \ref{results}. For the CHSH configuration, compare Eqs. \ref{hh1}, \ref{vv1}, \ref{hv1}, \ref{vh1}, \ref{Spsi}, \ref{Sphi} with Figs. \ref{CHSH_ID}, \ref{bars_CHSH} and Table \ref{results}. Clearly, these experiments verify the unique correlation signatures predicted for each entangled Bell state.
\begin{table}[t]
    \centering
    \begin{tabular}{| c | c | c | c | c |}
    \hline
                         &$\Psi^+ $&$ \Psi^-$ &$\Phi^+$ &$\Phi^-$ \\ \hline
    $\left\langle f+f^*\right\rangle$&0.96$\pm$0.01&-0.94$\pm$0.05 & -0.07$\pm$0.05 & 0.07$\pm$0.04\\ \hline
    $\left\langle d+d^*\right\rangle$&0.08$\pm$0.05&-0.07$\pm$0.05 & 0.90$\pm$0.04 & -0.90$\pm$0.01\\ \hline
    $\left\langle\mathcal{S}_{\psi}\right\rangle$&2.46$\pm$0.26&-2.51$\pm$0.35 & 0.04$\pm$0.36 & -0.01$\pm$0.23\\ \hline
    $\left\langle\mathcal{S}_{\phi}\right\rangle$&0.04$\pm$0.26&-0.05$\pm$0.35 & 2.57$\pm$0.36 & -2.66$\pm$0.23\\ \hline
    \end{tabular}
\caption{Experimental results for $\langle f+f^* \rangle$ and $\langle d+d^* \rangle$ and Bell parameters $\langle\mathcal{S}_{\psi}\rangle$ and $\langle\mathcal{S}_{\phi}\rangle$ using accidental corrected and normalized 100 second coincidence counts. These results demonstrate unique signatures for each Bell state. \label{results}}\end{table}

\section*{Conclusion}
We have reported a nonlocal interferometer capable of detecting entanglement and identifying Bell states statistically. We have shown that this is possible due to the NPI's unique correlation dependence on the anti-diagonal elements of the density matrix, which have distinct bounds for separable states and unique values for the four Bell states. We have introduced an NPI based CHSH-Bell test in which the ``local reality" is the photon polarization. The statistics resultant from the CHSH-Bell test were also shown to identify the Bell state. We have experimentally verified the NPI's sensitivity to the four Bell states in the primary configuration NPI$_{\pi\!/\!_4}$ and in the CHSH-Bell configuration NPI$_{\pi\!/\!_4}^\textrm{Bell}$. Our CHSH-Bell results have also exceeded the locality bound of $S\leq 2$ for each Bell state, as they should. Immediate quantum information applications of the NPI include experiments utilizing nonlocal interference to characterize entanglement. This includes statistical measurements of quantum channel noise, for example, in free-space optical communications. These diagnostic experiments often impose stringent timing constraints on the interrogated photons, which may be more easily realized using the ``same-time" phase-stable NPI design. The NPI also holds promise for quantum security applications, where measuring nonlocal interference is used as part of physical layer security \cite{6576339}. In conclusion, the capabilities of the nonlocal polarization interferometer in entanglement detection and statistical Bell state identification suggest application in current and future quantum information systems.

This work was supported by the Defense Threat Reduction Agency. This manuscript has been authored by UT-Battelle, LLC, under Contract No. DE-AC05-00OR22725 with the U.S. Department of Energy. We thank Lloyd Davis for comments on the experiment and equipment for its implementation.

\bibliography{bibliography}

\begin{thebibliography}{17}%
\makeatletter
\providecommand \@ifxundefined [1]{%
 \@ifx{#1\undefined}
}%
\providecommand \@ifnum [1]{%
 \ifnum #1\expandafter \@firstoftwo
 \else \expandafter \@secondoftwo
 \fi
}%
\providecommand \@ifx [1]{%
 \ifx #1\expandafter \@firstoftwo
 \else \expandafter \@secondoftwo
 \fi
}%
\providecommand \natexlab [1]{#1}%
\providecommand \enquote  [1]{``#1''}%
\providecommand \bibnamefont  [1]{#1}%
\providecommand \bibfnamefont [1]{#1}%
\providecommand \citenamefont [1]{#1}%
\providecommand \href@noop [0]{\@secondoftwo}%
\providecommand \href [0]{\begingroup \@sanitize@url \@href}%
\providecommand \@href[1]{\@@startlink{#1}\@@href}%
\providecommand \@@href[1]{\endgroup#1\@@endlink}%
\providecommand \@sanitize@url [0]{\catcode `\\12\catcode `\$12\catcode
  `\&12\catcode `\#12\catcode `\^12\catcode `\_12\catcode `\%12\relax}%
\providecommand \@@startlink[1]{}%
\providecommand \@@endlink[0]{}%
\providecommand \url  [0]{\begingroup\@sanitize@url \@url }%
\providecommand \@url [1]{\endgroup\@href {#1}{\urlprefix }}%
\providecommand \urlprefix  [0]{URL }%
\providecommand \Eprint [0]{\href }%
\providecommand \doibase [0]{http://dx.doi.org/}%
\providecommand \selectlanguage [0]{\@gobble}%
\providecommand \bibinfo  [0]{\@secondoftwo}%
\providecommand \bibfield  [0]{\@secondoftwo}%
\providecommand \translation [1]{[#1]}%
\providecommand \BibitemOpen [0]{}%
\providecommand \bibitemStop [0]{}%
\providecommand \bibitemNoStop [0]{.\EOS\space}%
\providecommand \EOS [0]{\spacefactor3000\relax}%
\providecommand \BibitemShut  [1]{\csname bibitem#1\endcsname}%
\let\auto@bib@innerbib\@empty
\bibitem [{\citenamefont {Pan}\ \emph {et~al.}(2012)\citenamefont {Pan},
  \citenamefont {Chen}, \citenamefont {Lu}, \citenamefont {Weinfurter},
  \citenamefont {Zeilinger},\ and\ \citenamefont {\ifmmode~\dot{Z}\else
  \.{Z}\fi{}ukowski}}]{multiphotonreview}%
  \BibitemOpen
  \bibfield  {author} {\bibinfo {author} {\bibfnamefont {J.-W.}\ \bibnamefont
  {Pan}}, \bibinfo {author} {\bibfnamefont {Z.-B.}\ \bibnamefont {Chen}},
  \bibinfo {author} {\bibfnamefont {C.-Y.}\ \bibnamefont {Lu}}, \bibinfo
  {author} {\bibfnamefont {H.}~\bibnamefont {Weinfurter}}, \bibinfo {author}
  {\bibfnamefont {A.}~\bibnamefont {Zeilinger}}, \ and\ \bibinfo {author}
  {\bibfnamefont {M.}~\bibnamefont {\ifmmode~\dot{Z}\else \.{Z}\fi{}ukowski}},\
  }\href {\doibase 10.1103/RevModPhys.84.777} {\bibfield  {journal} {\bibinfo
  {journal} {Rev. Mod. Phys.}\ }\textbf {\bibinfo {volume} {84}},\ \bibinfo
  {pages} {777} (\bibinfo {year} {2012})}\BibitemShut {NoStop}%
\bibitem [{\citenamefont {Ekert}(1991)}]{Ekert91}%
  \BibitemOpen
  \bibfield  {author} {\bibinfo {author} {\bibfnamefont {A.~K.}\ \bibnamefont
  {Ekert}},\ }\href {\doibase 10.1103/PhysRevLett.67.661} {\bibfield  {journal}
  {\bibinfo  {journal} {Phys. Rev. Lett.}\ }\textbf {\bibinfo {volume} {67}},\
  \bibinfo {pages} {661} (\bibinfo {year} {1991})}\BibitemShut {NoStop}%
\bibitem [{\citenamefont {Bennett}\ and\ \citenamefont
  {Wiesner}(1992)}]{superdense}%
  \BibitemOpen
  \bibfield  {author} {\bibinfo {author} {\bibfnamefont {C.~H.}\ \bibnamefont
  {Bennett}}\ and\ \bibinfo {author} {\bibfnamefont {S.~J.}\ \bibnamefont
  {Wiesner}},\ }\href {\doibase 10.1103/PhysRevLett.69.2881} {\bibfield
  {journal} {\bibinfo  {journal} {Phys. Rev. Lett.}\ }\textbf {\bibinfo
  {volume} {69}},\ \bibinfo {pages} {2881} (\bibinfo {year}
  {1992})}\BibitemShut {NoStop}%
\bibitem [{\citenamefont {Bennett}\ \emph {et~al.}(1993)\citenamefont
  {Bennett}, \citenamefont {Brassard}, \citenamefont {Cr\'epeau}, \citenamefont
  {Jozsa}, \citenamefont {Peres},\ and\ \citenamefont {Wootters}}]{teleport}%
  \BibitemOpen
  \bibfield  {author} {\bibinfo {author} {\bibfnamefont {C.~H.}\ \bibnamefont
  {Bennett}}, \bibinfo {author} {\bibfnamefont {G.}~\bibnamefont {Brassard}},
  \bibinfo {author} {\bibfnamefont {C.}~\bibnamefont {Cr\'epeau}}, \bibinfo
  {author} {\bibfnamefont {R.}~\bibnamefont {Jozsa}}, \bibinfo {author}
  {\bibfnamefont {A.}~\bibnamefont {Peres}}, \ and\ \bibinfo {author}
  {\bibfnamefont {W.~K.}\ \bibnamefont {Wootters}},\ }\href {\doibase
  10.1103/PhysRevLett.70.1895} {\bibfield  {journal} {\bibinfo  {journal}
  {Phys. Rev. Lett.}\ }\textbf {\bibinfo {volume} {70}},\ \bibinfo {pages}
  {1895} (\bibinfo {year} {1993})}\BibitemShut {NoStop}%
\bibitem [{\citenamefont {O'Brien}(2007)}]{qc}%
  \BibitemOpen
  \bibfield  {author} {\bibinfo {author} {\bibfnamefont {J.~L.}\ \bibnamefont
  {O'Brien}},\ }\href@noop {} {\bibfield  {journal} {\bibinfo  {journal}
  {Science}\ }\textbf {\bibinfo {volume} {318}},\ \bibinfo {pages} {1567}
  (\bibinfo {year} {2007})}\BibitemShut {NoStop}%
\bibitem [{\citenamefont {Horodecki}\ \emph {et~al.}(2009)\citenamefont
  {Horodecki}, \citenamefont {Horodecki}, \citenamefont {Horodecki},\ and\
  \citenamefont {Horodecki}}]{qe}%
  \BibitemOpen
  \bibfield  {author} {\bibinfo {author} {\bibfnamefont {R.}~\bibnamefont
  {Horodecki}}, \bibinfo {author} {\bibfnamefont {P.}~\bibnamefont
  {Horodecki}}, \bibinfo {author} {\bibfnamefont {M.}~\bibnamefont
  {Horodecki}}, \ and\ \bibinfo {author} {\bibfnamefont {K.}~\bibnamefont
  {Horodecki}},\ }\href {\doibase 10.1103/RevModPhys.81.865} {\bibfield
  {journal} {\bibinfo  {journal} {Rev. Mod. Phys.}\ }\textbf {\bibinfo {volume}
  {81}},\ \bibinfo {pages} {865} (\bibinfo {year} {2009})}\BibitemShut
  {NoStop}%
\bibitem [{\citenamefont {Gühne}\ and\ \citenamefont
  {Tóth}(2009)}]{entanglement_detection}%
  \BibitemOpen
  \bibfield  {author} {\bibinfo {author} {\bibfnamefont {O.}~\bibnamefont
  {Gühne}}\ and\ \bibinfo {author} {\bibfnamefont {G.}~\bibnamefont {Tóth}},\
  }\href {\doibase http://dx.doi.org/10.1016/j.physrep.2009.02.004} {\bibfield
  {journal} {\bibinfo  {journal} {Physics Reports}\ }\textbf {\bibinfo {volume}
  {474}},\ \bibinfo {pages} {1 } (\bibinfo {year} {2009})}\BibitemShut
  {NoStop}%
\bibitem [{\citenamefont {Clauser}\ \emph {et~al.}(1969)\citenamefont
  {Clauser}, \citenamefont {Horne}, \citenamefont {Shimony},\ and\
  \citenamefont {Holt}}]{CHSH}%
  \BibitemOpen
  \bibfield  {author} {\bibinfo {author} {\bibfnamefont {J.~F.}\ \bibnamefont
  {Clauser}}, \bibinfo {author} {\bibfnamefont {M.~A.}\ \bibnamefont {Horne}},
  \bibinfo {author} {\bibfnamefont {A.}~\bibnamefont {Shimony}}, \ and\
  \bibinfo {author} {\bibfnamefont {R.~A.}\ \bibnamefont {Holt}},\ }\href
  {\doibase 10.1103/PhysRevLett.23.880} {\bibfield  {journal} {\bibinfo
  {journal} {Phys. Rev. Lett.}\ }\textbf {\bibinfo {volume} {23}},\ \bibinfo
  {pages} {880} (\bibinfo {year} {1969})}\BibitemShut {NoStop}%
\bibitem [{\citenamefont {Vidal}\ and\ \citenamefont
  {Werner}(2002)}]{negativity}%
  \BibitemOpen
  \bibfield  {author} {\bibinfo {author} {\bibfnamefont {G.}~\bibnamefont
  {Vidal}}\ and\ \bibinfo {author} {\bibfnamefont {R.~F.}\ \bibnamefont
  {Werner}},\ }\href {\doibase 10.1103/PhysRevA.65.032314} {\bibfield
  {journal} {\bibinfo  {journal} {Phys. Rev. A}\ }\textbf {\bibinfo {volume}
  {65}},\ \bibinfo {pages} {032314} (\bibinfo {year} {2002})}\BibitemShut
  {NoStop}%
\bibitem [{\citenamefont {Horodecki}\ \emph {et~al.}(1996)\citenamefont
  {Horodecki}, \citenamefont {Horodecki},\ and\ \citenamefont
  {Horodecki}}]{Horodecki1996}%
  \BibitemOpen
  \bibfield  {author} {\bibinfo {author} {\bibfnamefont {M.}~\bibnamefont
  {Horodecki}}, \bibinfo {author} {\bibfnamefont {P.}~\bibnamefont
  {Horodecki}}, \ and\ \bibinfo {author} {\bibfnamefont {R.}~\bibnamefont
  {Horodecki}},\ }\href {\doibase
  http://dx.doi.org/10.1016/S0375-9601(96)00706-2} {\bibfield  {journal}
  {\bibinfo  {journal} {Physics Letters A}\ }\textbf {\bibinfo {volume}
  {223}},\ \bibinfo {pages} {1 } (\bibinfo {year} {1996})}\BibitemShut
  {NoStop}%
\bibitem [{\citenamefont {Terhal}(2000)}]{terhal2000}%
  \BibitemOpen
  \bibfield  {author} {\bibinfo {author} {\bibfnamefont {B.~M.}\ \bibnamefont
  {Terhal}},\ }\href {\doibase http://dx.doi.org/10.1016/S0375-9601(00)00401-1}
  {\bibfield  {journal} {\bibinfo  {journal} {Physics Letters A}\ }\textbf
  {\bibinfo {volume} {271}},\ \bibinfo {pages} {319 } (\bibinfo {year}
  {2000})}\BibitemShut {NoStop}%
\bibitem [{\citenamefont {Franson}(1989)}]{Franson89}%
  \BibitemOpen
  \bibfield  {author} {\bibinfo {author} {\bibfnamefont {J.~D.}\ \bibnamefont
  {Franson}},\ }\href {\doibase 10.1103/PhysRevLett.62.2205} {\bibfield
  {journal} {\bibinfo  {journal} {Phys. Rev. Lett.}\ }\textbf {\bibinfo
  {volume} {62}},\ \bibinfo {pages} {2205} (\bibinfo {year}
  {1989})}\BibitemShut {NoStop}%
\bibitem [{\citenamefont {Ng}\ \emph {et~al.}(2011)\citenamefont {Ng},
  \citenamefont {Gosal}, \citenamefont {Lamas-Linares},\ and\ \citenamefont
  {Kurtsiefer}}]{DDPM}%
  \BibitemOpen
  \bibfield  {author} {\bibinfo {author} {\bibfnamefont {T.~T.}\ \bibnamefont
  {Ng}}, \bibinfo {author} {\bibfnamefont {D.}~\bibnamefont {Gosal}}, \bibinfo
  {author} {\bibfnamefont {A.}~\bibnamefont {Lamas-Linares}}, \ and\ \bibinfo
  {author} {\bibfnamefont {C.}~\bibnamefont {Kurtsiefer}},\ }\href {\doibase
  http://dx.doi.org/10.1063/1.3514984} {\bibfield  {journal} {\bibinfo
  {journal} {Review of Scientific Instruments}\ }\textbf {\bibinfo {volume}
  {82}},\ \bibinfo {eid} {013106} (\bibinfo {year} {2011})}\BibitemShut
  {NoStop}%
\bibitem [{Note1()}]{Note1}%
  \BibitemOpen
  \bibinfo {note} {A non-degenerate signal and idler would destroy the
  polarization entanglement, since the photon energy would be associated with a
  specific polarization. This setup also leads to local two-photon interference
  when the photons both go to Alice or Bob's interferometers. A local effect
  observed is Hong-Ou-Mandel (HOM) interference which is maximized when the
  signal and idler are spectrally indistinguishable. The visibility of HOM
  interference was used to tune the signal and idler
  indistinguishability.}\BibitemShut {Stop}%
\bibitem [{Note2()}]{Note2}%
  \BibitemOpen
  \bibinfo {note} {See www.zedboard.com and www.xillibus.com.}\BibitemShut
  {Stop}%
\bibitem [{Note3()}]{Note3}%
  \BibitemOpen
  \bibinfo {note} {The signal and idler were made spectrally distinguishable by
  tuning the temperature of the PPKTP crystal. Distinguishability was verified
  by vanishing local HOM interference, see [14]. This distinguishability
  destroys the polarization entanglement.}\BibitemShut {Stop}%
\bibitem [{\citenamefont {Humble}(2013)}]{6576339}%
  \BibitemOpen
  \bibfield  {author} {\bibinfo {author} {\bibfnamefont {T.}~\bibnamefont
  {Humble}},\ }\href {\doibase 10.1109/MCOM.2013.6576339} {\bibfield  {journal}
  {\bibinfo  {journal} {Communications Magazine, IEEE}\ }\textbf {\bibinfo
  {volume} {51}},\  (\bibinfo {year} {2013})}\BibitemShut {NoStop}%
\end{thebibliography}%
\end{document}